\documentclass[thmsa,a4paper,oneside,12pt,onecolumn]{article}
\usepackage{sw20lart}

\input tcilatex
\QQQ{Language}{
American English
}

\begin{document}

\title{Vector Meson Propagator\\
and Baryon Current Conservation}
\author{S. S. Wu, H. X. Zhang, Y. J. Yao \\
%EndAName
Center for Theoretical Physics and Department of Physics \\
Jilin University, Changchun 130023 \\
People's Republic of China}
\date{}
\maketitle

\begin{abstract}
If baryons couple only with $\omega $-mesons, one found the baryon spectral
function may be negative. We show this unacceptable result is caused by the $%
k_\mu k_\nu $-terms in the $\omega $-meson propagator. Their contribution
may not vanish in approximate calculations which violate the baryon current
conservation. A rule is suggested, by which the calculated baryon spectral
function is well behaved.

PACS: 21.60. J$_z$, 21.65+f, 11.10Gh

Keywords: Vector meson propagator, baryon current conservation
\end{abstract}

Krein, Nielsen, Puff and Wilets (KNPW) [1] pointed out not long ago that in
a self-consistent (SC) Hartree-Fock (HF) calculation of the renormalized
baryon propagator for the case of zero-density the spectral function $%
A_R(\kappa )$ can be negative for some real values of $\kappa $, if baryons
couple only with $\omega $-mesons. They emphasized that this is
unacceptable. The spectral representation they considered is of the form 
\begin{equation}
G(k)=-\int_{-\infty }^\infty d\kappa A_R(\kappa )\frac{\gamma _\mu k_\mu
+i\kappa }{k^2+\kappa ^2-i\epsilon }.  \tag{1}
\end{equation}
Since $A_R(\kappa )$ represents the probability that a state of mass $%
|\kappa |$ is created, it must be non-negative. They suggested that it might
be due to the inadequacy of the HF approximation or the inconsistency of the
theory. In their calculation they have neglected all the terms proportional
to $k_\mu k_\nu $ in the $\omega $-meson propagator on the basis of the
baryon current conservation implied by the model for a rigorous calculation.
Though this is a generally accepted approximation [2] and indeed, such terms
need not be taken into account if the baryon current conserves, their
contribution in the SCHF approximation is not zero and has to be studied. It
also indicates that the SCHF approximation does not preserve the baryon
current conservation. The aim of this letter is to show that the baryon
spectral function becomes non-negative and is well behaved if an adequate
part of the $k_\mu k_\nu $-terms is included.

For a system of baryons coupling only with $\omega $-mesons, the Lagrangian
density has the form 
\begin{equation}
L=-\overline{\psi }(\gamma _\mu \partial _\mu +M)\psi -\frac 14F_{\mu \nu
}F_{\mu \nu }-\frac 12m_v^2A_\mu A_\mu +ig_v\overline{\psi }\gamma _\mu \psi
A_\mu +L_{CTC},  \tag{2}
\end{equation}
where $F_{\mu \nu }=\partial _\mu A_\nu -\partial _\nu A_\mu $, $\partial
_\mu \equiv \partial /\partial x_\mu $, $x_\mu =(\vec x,ix_0)$, $x^2=x_\mu
x_\mu =\vec x^2-x_0^{2\text{ }}$ with $x_0\equiv t$ and CTC means the
counterterm correction introduced for the purpose of renormalization. The
baryon and $\omega $-meson propagators are defined as 
\begin{equation}
G_{\alpha \beta }(x=x_1-x_2)=\left\langle T[\psi _\alpha (x_1)\overline{\psi 
}_\beta (x_2)]\right\rangle =\int \frac{d^4k}{(2\pi )^4}e^{ik_\rho x_\rho
}G_{\alpha \beta }(k),  \tag{3}
\end{equation}
\begin{equation}
D_{\mu \nu }(x)=\left\langle T[A_\mu (x_1)A_\nu (x_2)]\right\rangle =\int 
\frac{d^4k}{(2\pi )^4}e^{ik_\rho x_\rho }D_{\mu \nu }(k),  \tag{4}
\end{equation}
where $\left\langle 0\right\rangle \equiv \left\langle \Psi _0|0|\Psi
_0\right\rangle $ and $|\Psi _0\rangle $ denotes the exact ground state. The
dressed HF scheme is illustrated in Fig. 1, where $\overline{G}$ ($\overline{%
D}$) denotes an appropriate expression chosen for the calculation of the
baryon ($\omega $-meson) propagator in the self-energy. As is wellknown, the
relevant Dyson-Schwinger (DS) equations can be written as follows:

(a) for baryon 
\begin{equation}
G(k)=G^0(k)+G^0(k)\Sigma (k)G(k)=-[\gamma _\mu k_\mu -iM+\Sigma (k)]^{-1}, 
\tag{5}
\end{equation}
\begin{equation}
\Sigma (k)=g_v^2\int \frac{d^\tau q}{(2\pi )^4}\Gamma _\eta (k,q,k-q)%
\overline{G}(q)\overline{D}_{\eta \lambda }(k-q)\gamma _\lambda +\Sigma
_{CTC}(k);  \tag{6}
\end{equation}

(b) for $\omega $-meson 
\begin{equation}
D_{\mu \nu }(k)=D_{\mu \nu }^0(k)+D_{\mu \eta }^0(k)\Pi _{\eta \lambda
}(k)D_{\lambda \nu }(k),  \tag{7}
\end{equation}
\begin{equation}
\widehat{\Pi }_{\eta \lambda }(k)=-g_v^2\int \frac{d^\tau q}{(2\pi )^4}%
Tr[\Gamma _\eta (k+q,k,q)\overline{G}(k+q)\gamma _\lambda \overline{G}(q)], 
\tag{8}
\end{equation}
where a caret indicates that the quantity is not yet renormalized, $\tau
=4-\delta $ ($\delta \rightarrow 0^{+}$) and in Eq. (5) the Feynman
prescription $M\rightarrow M-i\epsilon $ is understood. Since the baryon
current is conserved, we have 
\begin{equation}
\widehat{\Pi }_{\eta \lambda }(k)=(\delta _{\eta \lambda }-\frac{k_\eta
k_\lambda }{k^2})\widehat{\Pi }_v(k),  \tag{9}
\end{equation}
which gives $\widehat{\Pi }_v(k)=\frac 13\sum_\eta \widehat{\Pi }_{\eta \eta
}(k)$. Eq. (9) shows that the renormalized $\Pi _{\eta \lambda }$ can be
obtained from 
\begin{equation}
\Pi _v(k)=\widehat{\Pi }_v(k)+\Pi _{CTC}(k).  \tag{10}
\end{equation}
The renormalized counterterms will be fixed by the following conditions 
\begin{equation}
\Sigma (k)|_{\gamma _\mu k_\mu =iM_t}=0;\qquad \frac{\partial \Sigma (k)}{%
\partial (\gamma _\mu k_\mu )}|_{\gamma _\mu k_\mu =iM_t}=0,  \tag{11a}
\end{equation}
\begin{equation}
\Pi _v(k)|_{k^2=0}=0;\qquad \frac{\partial \Pi _v(k)}{\partial k^2}%
|_{k^2=0}=0,  \tag{11b}
\end{equation}
where $M_t$ is the true baryon mass. Following [3], we shall use the
following notations to distinguish different approximations: scheme P, $%
\overline{G}=-[\gamma _\mu k_\mu -iM_t]^{-1}$, $\overline{D}_{\eta \lambda
}=D_{\eta \lambda }^0$; scheme BP, $\overline{G}=G$, $\overline{D}_{\eta
\lambda }=D_{\eta \lambda }^0$ and scheme FSC, $\overline{G}=G$, $\overline{D%
}_{\eta \lambda }=D_{\eta \lambda }$. It is seen that in scheme BP Eq. (5)
has to be solved self-consistently [1, 4-6], while in scheme FSC [7, 3] one
should consider a self-consistent coupled set of renormalized DS equations
(5-10). As emphasized in [7], the study of an ansatz consistent with the
Ward-Takahashi identity may provide a way to preserve the baryon current
conservation. Though it does not hold in the SCHF approximation, we shall
assume that Eq. (9) is valid approximately. We note that $D_{\mu \nu }$ can
be written as [8, 9] 
\begin{equation}
D_{\mu \nu }(k)=(\delta _{\mu \nu }-\frac{k_\mu k_\nu }{k^2})\Delta _v(k)-i%
\frac{k_\mu k_\nu }{k^2(m_v^2+\delta m_v^2)},  \tag{12}
\end{equation}
\begin{equation}
\Delta _v(k)=\Delta _v^0(k)+\Delta _v^0(k)\Pi _v(k)\Delta
_v(k)=-i[k^2+m_v^2+i\Pi _v(k)-i\epsilon ]^{-1},  \tag{13}
\end{equation}
where $\delta m_v^2$ is the mass counterterm for the $\omega $-meson. As is
wellknown [10], a properly normalized spectral representation for the baryon
propagator equivalent to Eq. (1) can be written in the form: 
\begin{equation}
\widetilde{G}(k)=-Z_2\frac{\gamma _\mu k_\mu +iM_t}{k^2+M_t^2-i\epsilon }%
-\dint_{m_1^2}^\infty d\sigma ^2\frac{\gamma _\mu k_\mu \alpha (-\sigma
^2,Z_2)+iM_t\beta (-\sigma ^2,Z_2)}{k^2+\sigma ^2-i\epsilon },  \tag{14}
\end{equation}
where $m_1=M_t+m_v$ is the threshold of the continuum spectrum, 
\begin{equation}
Z_2+\dint_{m_1^2}^\infty d\sigma ^2\alpha (-\sigma ^2,Z_2)=1  \tag{15}
\end{equation}
and the spectral functions $\alpha $ and $\beta $ should have the following
properties:

\begin{equation}
\begin{array}{l}
\text{(a).}\;\text{they\thinspace are\thinspace all\thinspace real,} \\ 
\text{(b).}\;\alpha (-\sigma ^2,Z_2)\geq 0, \\ 
\text{(c).}\;\sigma \alpha (-\sigma ^2,Z_2)-M_t\beta (-\sigma ^2,Z_2)\geq 0.
\end{array}
\tag{16}
\end{equation}
For the $\omega $-meson propagator we have 
\begin{equation}
\widetilde{\Delta }_v(k)=Z_v\frac{-i}{k^2+\widehat{m}_v^2-i\epsilon }%
-i\dint_{th}^\infty dm^2\frac{\rho _v(-m^2,Z_v)}{k^2+m^2-i\epsilon }, 
\tag{17a}
\end{equation}
\begin{equation}
Z_v+\dint_{th}^\infty dm^2\rho _v(-m^2,Z_v)=1.  \tag{17b}
\end{equation}
In Eq. (17) $\widehat{m}_v$ denotes the true $\omega $-meson mass and $%
th=(2M_t)^2$ the threshold of the continuum. Since $G(k)$ and $\widetilde{G}%
(k)$ as well as $\Delta _v(k)$ and $\widetilde{\Delta }_v(k)$ are normalized
differently, we should write $\widetilde{G}(k)=ZG(k)$ and $\widetilde{\Delta 
}_v(k)=Y_v\Delta _v(k)$. It is easily seen that $Z_2=ZZ_t$ and $Z_v=Y_vR_v$,
where ($-Z_t$) and ($-iR_v$) are the residues of $G(k)$ and $\Delta _v(k)$
at the poles $\gamma _\mu k_\mu =iM_t$ and $k^2=-\widehat{m}_v^2$,
respectively. In the case of zero-density one may write 
\begin{equation}
\Sigma (k)=\gamma _\mu k_\mu a(k^2)-iMb(k^2).  \tag{18}
\end{equation}
We note that under the on-shell condition (11a) one has $Z_t=1$ and $M=M_t$
[6]. Neglecting all the $k_\mu k_\nu $ terms in Eq. (12) and substituting $%
\overline{D}_{\mu \nu }=\delta _{\mu \nu }\Delta _v=\delta _{\mu \nu
}Y_v^{-1}\widetilde{\Delta }_v$ as well as $\overline{G}=Z^{-1}\widetilde{G}$
in Eqs. (6) and (8), an explicit coupled set of renormalized equations for
the determination of ($a,b$), ($\alpha ,\beta $), $\Pi _v$ and $\rho _v$ has
been given in [3]. Since the expressions for the rest of the equations
considered here are the same as in [3], for space saving and for convenience
of discussion only a part of them will be written down as follows: 
\begin{equation}
a(k^2)=a_\delta (k^2)+a_\Delta (k^2);\qquad b(k^2)=b_\delta (k^2)+b_\Delta
(k^2)  \tag{19}
\end{equation}
\begin{equation}
a_\delta (k^2)=\frac{g_v^2}{8\pi ^2}\diint_0^\infty d\sigma
^2dm^2\dint_0^1dxf_\alpha (-\sigma ^2)h_v(-m^2)(1-x)\ln \frac{K^2(-M_t^2)}{%
K^2(k^2)}+c_\delta ,  \tag{19a}
\end{equation}
\begin{equation}
b_\delta (k^2)=\frac{g_v^2}{4\pi ^2}\diint_0^\infty d\sigma
^2dm^2\dint_0^1dxf_\beta (-\sigma ^2)h_v(-m^2)\ln \frac{K^2(-M_t^2)}{K^2(k^2)%
}+c_\delta ,  \tag{19b}
\end{equation}
\begin{equation}
f_\gamma (-\sigma ^2)=\delta (\sigma ^2-M_t^2)+\theta (\sigma
^2-m_1^2)\gamma (-\sigma ^2),\quad \text{(}\gamma =\alpha \text{ or }\beta 
\text{)}  \tag{19c}
\end{equation}
\begin{equation}
h_v(-m^2)=R_v\delta (m^2-\widehat{m}_v^2)+\theta (m^2-th)\rho _v(-m^2), 
\tag{19d}
\end{equation}
\begin{equation}
K^2(k^2)\equiv K^2(x,k^2,\sigma ^2,m^2)=x(1-x)k^2+x\sigma ^2+(1-x)m^2, 
\tag{19e}
\end{equation}
where $\gamma (-\sigma ^2)=Z^{-1}\gamma (-\sigma ^2,Z_2)$ and $\rho
_v(-m^2)=Y_v^{-1}\rho _v(-m^2,Z_v)$. In addition, we have 
\begin{equation}
\alpha (k^2)=\frac 1\pi Im\frac{1+a(k^2)}{D(k^2)};\qquad \beta (k^2)=\frac
1\pi Im\frac{1+b(k^2)}{D(k^2)},  \tag{20a}
\end{equation}
\begin{equation}
D(k^2)=[1+a(k^2)]^2k^2+[1+b(k^2)]^2M_t^2.  \tag{20b}
\end{equation}
In Eq. (19) the $\Delta $-terms are contributed by the second $k_\mu k_\nu $%
-term in the round brackets of Eq. (12). They are given by 
\begin{equation}
\begin{array}{c}
a_\Delta (k^2)=-\frac{g_v^2}{16\pi ^2}\diint_0^\infty d\sigma
^2dm^2\dint_0^1dx\dint_0^{1-x}dyf_\alpha (-\sigma ^2)h_v(-m^2)\times \\ 
\{(1+3x)\ln \frac{L^2(-M_t^2)}{L^2(k^2)}-x^2(1-x)[\frac{k^2}{L^2(k^2)}+\frac{%
M_t^2}{L^2(-M_t^2)}]\}+c_\Delta
\end{array}
,  \tag{21a}
\end{equation}
\begin{equation}
\begin{array}{c}
b_\Delta (k^2)=-\frac{g_v^2}{16\pi ^2}\diint_0^\infty d\sigma
^2dm^2\dint_0^1dx\dint_0^{1-x}dyf_\beta (-\sigma ^2)h_v(-m^2)\times \\ 
\{2\ln \frac{L^2(-M_t^2)}{L^2(k^2)}+x^2[\frac{k^2}{L^2(k^2)}+\frac{M_t^2}{%
L^2(-M_t^2)}]\}+c_\Delta
\end{array}
,  \tag{21b}
\end{equation}
\begin{equation}
L^2(k^2)\equiv L^2(x,y,k^2,\sigma ^2,m^2)=x(1-x)k^2+x\sigma ^2+ym^2, 
\tag{21c}
\end{equation}
where in Eqs. (19) and (21) we have $c_\kappa =2M_t^2[a_\kappa ^{\prime
}(-M_t^2)-b_\kappa ^{\prime }(-M_t^2)]$, $\kappa =\delta $ or $\Delta $, and 
$f^{\prime }(k^2)=df(k^2)/dk^2$.

As shown in Eq. (20), in order to find $\alpha $ and $\beta $ we need to
know the imaginary parts of $a$ and $b$. Eqs. (19) and (21) show that $a$
and $b$ will be complex if either $K^2(k^2)$ or $L^2(k^2)$ becomes negative.
Define $\theta _\kappa \equiv \theta _\kappa (x,\sigma ^2,m^2,k^2)=0$ if $%
K^2(k^2)>0$ and $\theta _\kappa =1$ if $K^2(k^2)<0$. We may write $\ln
K^2(k^2)=\ln |K^2(k^2)|\pm i\theta _\kappa (\pi +2n\pi )$. Even if we set $%
n=0$, $\ln K^2$ is still indeterminate in case $\theta _\kappa =1$. Let the
sign before $i\pi $ be denoted by s$\ln K^2$. Obviously the same remark also
applies to $\ln L^2(k^2)$. We shall make the stipulation that s$\ln K^2$ and
s$\ln L^2$ are to be so chosen that Eq. (16) should be satisfied. Since $a$
and $b$ are derived from the same righthand side of Eq. (6), their s$\ln K^2$
(s$\ln L^2$) should be the same. According to [3], the above coupled set of
equations can be solved easily by the method of iteration. We have solved it
for schemes BP and FSC. In the following we shall choose $M_t=4.7585fm^{-1}$%
, $\widehat{m}_v/M_t=0.83387$ and $\overline{g}_v^2=g_v^2/8\pi ^2=1.3685$.
If we neglect all the $k_\mu k_\nu $-term in Eq. (12), it is impossible to
make $\alpha (k^2)$ non-negative by an appropriate choice of s$\ln K^2$ and s%
$\ln L^2$. This is unacceptable and confirms the result of KNPW. As is
wellknown, the last term in Eq. (12) is not renormalizable and the model
becomes renormalizable only, because the $\omega $-meson couples to the
conserved baryon current so that this term may be neglected. Thus we have
calculated the additional contribution of the second term in the round
brackets of Eq. (12) according to Eq. (21). The numerical results for s$\ln
K^2=-$ and s$\ln L^2=+$ are shown in Fig. 2a. They are the best results we
can find and satisfy all the requirements of Eq. (16). However, they are
still physically unacceptable, because the threshold is wrong. It should be $%
-(M_t+\widehat{m}_v)^2$ rather than $-M_t^2$. This implies that the results
contain a contribution from an additional zero-mass meson, which does not
exist in our model. Clearly the latter is caused by the fact that $\frac{%
k_\mu k_\nu }{k^2}\Delta _v$ contains an additional zero-mass pole, because $%
k^2=\vec k^2-k_0^2=0$ does not ensure that $k_\mu $ and $k_\nu $ are both
zero. Since Eq. (12) is formally rigorous and $D_{\mu \nu }$ should contain
no zero-mass pole, the renormalized contribution of the latter must be
cancelled by a finite part arising from the last term. Indeed, this can also
be seen from the zero-order approximation to $D_{\mu \nu }$, which reads 
\begin{equation}
D_{\mu \nu }^0(k)=(\delta _{\mu \nu }-\frac{k_\mu k_\nu }{k^2})\Delta
_v^0(k)-i\frac{k_\mu k_\nu }{k^2m_v^2}=(\delta _{\mu \nu }+\frac{k_\mu k_\nu 
}{m_v^2})\Delta _v^0(k),  \tag{22}
\end{equation}
where $\Delta _v^0(k)=-i[k^2+m_v^2-i\epsilon ]^{-1}$. Thus, if we neglect
the last term in Eq. (12), $D_{\mu \nu }^0$ will contain a false zero-mass
pole. Eq. (22) indicates that this last term makes the pole structure of $%
D_{\mu \nu }^0$ correct, though unrenormalizable. Our task is then to
eliminate the contribution of the zero-mass meson from Eq. (21). This can be
done unambiguously. We note that the integrals over the parameter $y$ in Eq.
(21) can be performed exactly and in an elementary way. The expressions are,
however, too lengthy to write down here. Let the curved brackets in Eqs.
(21a and b) be denoted by $Y_c(y)$ with $c=a$ and $b$, respectively. We may
express the integration over $y$ as 
\begin{equation}
\int_0^{1-x}dyY_c(y)=Z_c(1-x)-Z_c(0).  \tag{23}
\end{equation}
Set $L^2(y)\equiv L^2(x,y,k^2,\sigma ^2,m^2)$. Eq. (19e) and (21c) show 
\begin{equation}
L^2(y=0)=K^2(x,k^2,\sigma ^2,0);\quad L^2(y=1-x)=K^2(x,k^2,\sigma ^2,m^2). 
\tag{24}
\end{equation}
Form Eqs. (21), (23) and (24) one easily finds that the contributions to the
wrong threshold come exclusively from $Z_c(0)$. Substituting Eq. (23) in Eq.
(21), discarding the contribution of $Z_c(0)$ and using Eq. (19), we have
again solved the coupled set of equations. Fig. 2b depicts our numerical
results for $\alpha (k^2)$ and $\beta (k^2)$ with the choice s$\ln K^2=-$
and s$\ln L^2=-$. They satisfy all the requirements of Eq. (16) and have a
correct threshold. Thus they are now acceptable. As displayed in Fig. 2, the
effect of self-consistency on the baryon spectral function is perceptible,
though the results of schemes BP and FSC are almost the same. The latter
asserts, in agreement with [3], that there is no need to include the meson
propagator in the self-consistency requirement. Clearly the fulfillment of
Eq. (16) also implies that $A_R(\kappa )$ will be non-negative. Our results
show if in an approximation the contribution of the renormalizable $k_\mu
k_\nu $-term in $D_{\mu \nu \text{ }}$is not zero, it must be taken account
of, otherwise the result may be qualitatively wrong. According to our above
results the rule for the calculation of this additional contribution may be
stated as follows: for $c=a$ or $b$, $Z_c(0)$ and the last term in Eq. (12)
should both be discarded, because the contributions of the former and of a
finite part from the latter cancel each other, while for the sake of
renormalization the infinite nonrenormalizable part need not be considered
on the basis of baryon current conservation. The latter can be understood as
follows. Let $A=A_0+A_r+A_\infty $ be a term calculated in the SCHF
approximation, where the subscripts $0$, $r$ and $\infty $ denote the
contributions from the first, second (renormalizable) and third
(nonrenormalizable) term in Eq. (12), respectively. Since the model
possesses the baryon current conservation, if $A$ violates it, there must
exist another term $B$ in the theory such that $A+B$ preserves it. Thus, $%
A_r+B_r$ and $A_\infty +B_\infty $ can be neglected. Since the infinity is
an indefinite quantity and it always contains an undefined finite part
(indicated by $f$), we may write $A_\infty +B_\infty =A_\infty ^{\prime
}+B_\infty ^{\prime }+A_f+B_f$. Thus, we are allowed to throw away $A_\infty
^{\prime }+B_\infty ^{\prime }$, if an adequate rule is prescribed to retain 
$A_f+B_f$. If only $A$ is calculated and it can already yield a quite good
approximation, clearly $A_r$ and $A_f$ must be taken into account except
when they are very small.

Redmond [11] suggested that one may use the spectral representation to
eliminate the ghost poles. Here we would like to point out that there is a
simple and direct way to derive a ghost-free representation for the Dyson
solution by the spectral representation. From Eqs. (5), (14) and 
\begin{equation}
G(k)=-D^{-1}[\gamma _\mu k_\mu \left( 1+a\right) +iM_t\left( 1+b\right)
]=Z^{-1}\widetilde{G}(k),  \tag{25}
\end{equation}
where $D$ is given by Eq. (20b), we easily find 
\begin{equation}
1+a=F(k^2)^{-1}F_\alpha (k^2);\qquad 1+b=F(k^2)^{-1}F_\beta (k^2),  \tag{26a}
\end{equation}
\begin{equation}
F(k^2)=k^2F_\alpha (k^2)^2+M_t^2F_\beta (k^2)^2  \tag{26b}
\end{equation}
with $F_\gamma \left( k^2\right) =\int_0^\infty d\sigma ^2f_\gamma (-\sigma
^2)(k^2+\sigma ^2-i\epsilon )^{-1}$ ($\gamma =\alpha $ or $\beta $). It is
seen that if $\alpha (k^2)$ and $\beta (k^2)$ are known, so are $F_\alpha
(k^2)$ and $F_\beta (k^2)$. Let us use a subscript $R$ to indicate the
results obtained from Eq. (26). Since $a_{R\text{ }}$ and $b_{R\text{ }}$
are direct consequences of the spectral representation $\widetilde{G}(k)$, $%
G_R(k)$ deduced from these $a_{R\text{ }}$ and $b_{R\text{ }}$ by means of
Eq. (5) satisfies Eq.(25) rigorously and is thus ghost-free, because $%
\widetilde{G}(k)$ is well behaved and possesses the Herglotz property [12].
Further one observes that $G_R(k)$ obviously satisfies the following Dyson
equation: 
\begin{equation}
G_R(k)=G^0(k)+G^0(k)\Sigma _R(k)G_R(k),  \tag{27}
\end{equation}
with $G^0(k)=-[\gamma _\mu k_\mu -iM_t]^{-1}$ and $\Sigma _R(k)=\gamma _\mu
k_\mu a_R(k^2)-iM_tb_R(k^2)$. Let us insert $\alpha $ and $\beta $ found
from Eq.(20) with $a=a_\delta +a_\Delta $ and $b=b_\delta +b_\Delta $
calculated according to Eqs.(19) and (21) in $F_\alpha $ and $F_\beta $.
From Eq. (26) we obtain $a_R$ and $b_R$. We note that the set ($a_R$, $b_R$)
is different from ($a$, $b$), though the former is deduced from the latter.
In fact, if we substitute ($a$, $b$) into Eq. (25), the latter no longer
holds, because $G(k)$ found in this way possesses ghost poles, which are $%
-100.4468\pm i55.2553fm^{-2}$. The reason is clear. To calculate $a$ and $b$
we need renormalization, which, however, does not ensure that Eq. (25)
holds. In Fig. 3 we have compared ($a_R,b_R$) with ($a,b$). They have
different asymptotic behavior and differ widely, though the same $\alpha $
and $\beta $ can be derived from them by Eq. (20). Our numerical result for $%
\rho _v\left( k^2\right) $ is well behaved and is qualitatively similar to $%
\rho _v\left( k^2\right) $ found for the $\sigma -\omega $ model in [3].
Using Eqs. (13, 17) and the relation $\Delta _v(k^2)=Y_v^{-1}\widetilde{%
\Delta }_v(k^2)$, we can also derive an analytic expression for $\Pi _v(k^2)$%
, which will yield a ghost-free $\Delta _v\left( k^2\right) $ by means of
Eq. (13). However, for lack of spacing both $\rho _v$ and $\Pi _v$ will not
be discussed here. As shown in Refs. [1, 3, 4, 7], if in addition to the $%
\omega $-meson, other mesons, for instance $\pi $, $\sigma $ and chiral $\pi
-\sigma $ (linear $\sigma $-model), are considered, the baryon spectral
functions in the self-consistent HF approximation are regular, even though
the $k_\mu k_\nu $-terms in Eq. (12) are neglected. The question what their
effects are in these more complicated cases is being studied.

The work is supported in part by the National Natural Science Foundation of
China and the Foundation of Chinese Education Ministry.

\newpage\

Fig. 1 on P$_2$, Fig. 2 on P$_6$, Fig. 3 on P$_7$.

\end{document}